\documentclass[pra,%
 reprint,
 amsmath,amssymb,
 aps,
]{revtex4-2}

\usepackage{graphicx}% Include figure files
\usepackage{dcolumn}% Align table columns on decimal point
\usepackage{bm}% bold math
\usepackage{hyperref}% add hypertext capabilities
\usepackage{physics}
\usepackage{algorithm}
\usepackage{algpseudocode}
\usepackage{amsmath,amsfonts,amssymb}
\usepackage{natbib}
\usepackage{multirow}
\usepackage{adjustbox}
\usepackage{lmodern}
\usepackage{xcolor}
\usepackage{amsmath,graphics}
\usepackage{ragged2e}
\usepackage{subcaption}

\usepackage{listings}
\begin{document}

\preprint{APS/123-QED}

\title{Distillation of continuous variable qudits from single photon sources: A cascaded approach}% Force line breaks with \\ Quantum state engineering via cascaded single photon catalysis

\author{Devibala Esakkimuthu}
\email{devibalaesakkimuthu@gmail.com}
\author{Basherrudin Mahmud Ahmed Abduljaffer}%
 \email{abmahmed.physics@mkuniversity.ac.in}
\affiliation{Department of Theoretical Physics, School of Physics,\\
Madurai Kamaraj Univeristy, Madurai-625021, Tamil Nadu, India.}

\date{\today}

\begin{abstract}
Creation of high fidelity photonic quantum states in the continuous variable regime is indispensable for the implementation of quantum technologies universally. However, this is a challenging task as it requires higher nonlinearity or larger Fock states. In this article, we surmount this necessity by using a linear optical setup with a cascaded arrangement of beam splitters that relies solely on single photon sources and single photon detectors to tailor desired single mode nonclassical states. To show the utility of this setup, we demonstrate the generation of displaced higher photon states with unit fidelity and the family of Schrodinger cat states above $98\%$ fidelity. In addition, we manifest the generation of GKP resource states, such as ON states and weak cubic phase states with $99\%$ fidelity. Creating such a variety of important states in this single setup is made feasible by stating the output in the form of displaced qudits. This figure of merit facilitates efficient identification and optimization of input parameters required to generate the target single mode quantum states. We also account for the experimental imperfections by incorporating detector inefficiencies and non-unit single photon sources. This cascaded setup will assist the experimentalists to explore the feasible creation of target states using currently available resources, such as single photon sources and single photon detectors.
\end{abstract}

\maketitle

\section{Introduction}
Quantum information processing has shown unmatched advantages over its classical counterpart in terms of efficiency and speed \cite{Nielsen2010}. Harnessing the quantum advantage requires the encoding, processing, and decoding of information in the form of quantum states. Various physical systems offer encoding of information in quantum systems. Among them, quantum optical systems have emerged as front runners due to their availability and agility \cite{Dellanno2006, Lvovsky2009}. In contrast to the classical case, where information is encoded in binary format, quantum mechanics offers multiple possibilities. Depending on whether the information is encoded in a finite or infinite dimensions, the physical systems are classified as discrete variable (DV) or continuous variable (CV) systems \cite{Andersen2015}. Generating quantum states in the CV regime is more accessible and has inherent advantages owing to currently available technologies \cite{Braunstein2005, Andersen2010}.

Generally, creating a CV nonclassical state or an arbitrary quantum state mandates greater nonlinearity. Attaining such level of nonlinearity is laborious, as the interaction Hamiltonian for higher-order polynomials is quite weak \cite{Corona2011, Senellart2017}. Meanwhile, one can also create a variant of quantum states using conditional measurements (CM) that can overcome the constraint of requiring higher nonlinearity. Prominent examples in the production of nonclassical states employing CM are Schrodinger cat like states \cite{Dakna1998c}, photon added states \cite{Dakna1998a}, and advancements in quantum state engineering \cite{Dakna1998b, Dakna1999a, Dakna1999b}. Even though the conditional methods are probabilistic, they deliver sufficient nonlinearity to produce effective nonclassical states by only using linear optical elements \cite{Lapaire2003}. Further the enhancement of nonclassicality utilizing CM, such as quantum entanglement \cite{Bartley2015, Zubairy2017, Yuan2019, Liu2022}, quantum teleportation \cite{Xu2015}, phase estimation \cite{Kumar2022, kumar2023, Manali2024}, squeezing distillation \cite{Kumar2024}, and increasing quantum key distribution distances \cite{Kumar2024a}, are achieved. While these states are advantageous, the existing CM protocols involve states with higher Fock components, but extending the input Fock states beyond a single photon is experimentally difficult task.

In this work, we overcome the restraints of using higher nonlinearities and larger Fock states by replacing them with linear optical components and single photon sources. Our setup consists a cascaded configuration of beam splitters with a coherent state as main input and performing single photon catalysis processes. Our aim is to create larger nonclassical resources such as higher Fock states and squeezed states with the utilization of minimal nonclassical resources i.e. single photon sources. In 2002, Lvovsky et al. introduced the quantum optical catalysis (QOC) technique, a subtype of conditional measurement \cite{Lvovsky2002}. This method involves using a single photon at one input port and a coherent state at the other, then detecting a single photon at one output port. This process creates a superposition of vacuum and single photon states. The injection and detection of single photon, which acts as a catalyst to generate a new quantum state, is termed as quantum optical catalysis. Employing QOC, several protocols using coherent state as one of the inputs have emerged namely, hole burning \cite{Baseia2004}, Fock filtration \cite{Sanaka2005}, orthogonalization of coherent state pairs \cite{Kruse2017}, multi-photon catalysis \cite{Bartley2012, Zubairy2016}.

As an extension of QOC, Sanaka et al, investigated the extraction of photon-number Fock states from coherent states by using a cascaded arrangement of beam splitters with single photon catalysis \cite{Sanaka2005}. They reported the creation of $\ket{5}$ with a maximum $80\%$ fidelity. The main draw back of this approach is the difficulty of removing the vacuum state in the generated states. Subsequently, Escher et al. developed a technique for generating higher Fock states using the cascaded arrangement of BS \cite{Escher2005}. Their approach involved introducing single photon states $\ket{1}$ into each input port of the beam splitters while configuring the detection system to register vacuum states $\ket{0}$. This method enabled the production of various higher Fock states, in particular $\ket{2}$, $\ket{5}$, and $\ket{10}$, all with perfect unit fidelity. However, using this method, the creation of superposition states is not possible, as it incorporates only DV. In 2019, Eaton et al. investigated the generation of superposed states using a cascaded arrangement of BS with coherent state and single photons as input \cite{Miller2019}. By changing the photon counting in the detector, they reported the creation of a superposition of squeezed vacuum state, squeezed cat states and GKP states. 

Even though the realization of several nonclassical states employing the cascaded setup has been reported, the full potential of this approach for generating arbitrary quantum states remains unexplored. It is necessary to fully characterize the system being used, which in itself is an demanding task. In this study, we subdue the difficulty of removing the vacuum component in the generation of higher photon states by expressing the cascaded output in the terms of displaced qudits (DQ). This figure of merit direct a path for target state manipulation and facilitates the identification of the appropriate input parameters. Since qudits are superposition of finite d-dimensional number states, this framework naturally enables the creation of superposition states as the result of taking coherent state as main input. A key feature of our approach is that it relies solely single photon detectors, yet it permits the creation of exotic nonclassical states are feasible. The free parameters to tune in our setup are coherent amplitude and reflectivity of beam splitters in the cascaded configuration.  

As a demonstration of our method, we generate a diverse  nonclassical states, including higher photon states, family of Schrodinger cat states, optimal squeezed states and GKP resource states. The higher photon states are produced with unit fidelity by concentrating the nonclassicality of all the input single photons; specifically, using $n-$ single photon sources enables the preparation of the $\ket{n}$ photon states. In addition, we optimize the input parameters of our setup to create the Schrodinger even/odd cat states, three headed cat states and compass states with fidelities exceeding $98\%$. By altering the superposition amplitudes of DQ, we further construct GKP resource states, such as the ON state and the weak cubic phase state. To the end, we present a general protocol for generating arbitrary quantum states using our cascaded approach. Although the resulting states are produced probabilistically, they offer significant advantages in terms of resource utilization. The primary limitation of the approach is the low success probability, which arises from the use of multiple beam splitters in the cascade. 

This article is arranged as follows: Sect. \ref{sect.DQ}, introduces the cascaded beam splitter setup and its output in displaced qudits form. In Sect. \ref{sect.Fock}, we discuss the generation of displaced Fock states using our protocol. The creation of  the family of Schrodinger cat states including three headed cat states and compass states, from the finite superposition of number states are described in \ref{sect.FSCS}. In Sect. \ref{sect.FSNS}, we show the possibilities of creating optimal squeezed states in finite superposition of number states and GKP resource states. Sect. \ref{sec:arbitrary}, presents an algorithm for the arbitrary quantum states generation. The realistic generation of the desired states with photon source inefficiency and photon detector inefficiency are discussed in Sect. \ref{sect.RQOC}. In Sect. \ref{sect.Sum}, we conclude the article with a summary of results and future directions.

\section{Displaced Qudits}
\label{sect.DQ}
%%%%%%%%%%%%%%%%%%%%%%%%%%%%%%%%%%%%%%%%%%
Our protocol involves single photon catalysis using a series of $l-$cascaded beam splitters. We start with a coherent state as the primary input. Each beam splitter undergoes a single photon catalysis process, where a single photon is injected and detected at specific ports to generate a new quantum state. The output from one beam splitter is then used as the input for the next beam splitter in the sequence, creating a cascaded system for quantum state manipulation.

The beam splitters operation can be described mathematically by unitary operator $\hat{U}_{l} = \exp\{\theta_l(\hat{a}^{\dagger}\hat{b}-\hat{a}\hat{b}^{\dagger})\}$. Here, $\theta_l=\cos^{-1}(\sqrt{R_l})$ is taken under the condition of $R_l+T_l=1$ and $R_l$ is the reflectivity of the respective BS. The output of the first beam splitter with reflectivity $R_1$ can be calculated by the steps;
\begin{align*}
    \ket{\psi}_1 = & \, _{b}\bra{1} \, \hat{U}_{1}\ket{\alpha}_a\ket{1}_b, \\
    = & \, _{b}\bra{1} \, \hat{U}_{1} \sum_{k=0}^{\infty} e^{-\abs{\alpha}^2} \frac{\alpha^k}{\sqrt{k!}} \ket{k}_a\ket{1}_b, \\
    = & \, e^{-\abs{\alpha}^2/2} \sum_{k=0}^{\infty} \frac{\alpha^k}{\sqrt{k!}} \, _{b}\bra{1} \, \hat{U}_{1} (\hat{a}^{\dagger})^k \ket{0}_a \,  \hat{b}^{\dagger} \ket{0}_b.
\end{align*}
The unitary operator $\hat{U}_{l}$,  transforms the two modes of field operators $\hat{a}$ and $\hat{b}$ for the input and ancilla respectively as per \cite{Holger2002}, 
\begin{align*}
    \hat{a} & \Rightarrow \sqrt{R_l}\, \hat{a}+i\sqrt{1-R_l}\,\hat{b}, \\
    \hat{b} & \Rightarrow i\sqrt{1-R_l}\, \hat{a}+i\sqrt{R_l}\,\hat{b}.
\end{align*}
After a certain amount of calculation, the expected output of $BS_1$ following the single photon catalysis process can be expressed as: 
\begin{align*}
   \ket{\psi}_1 = & \, \hbox{N}_1 \, e^{-\abs{\alpha}^2(1-R_1)/{2}} \big[\sqrt{R_1} - (1-R_1) \, \alpha \,\hat{a}^{\dagger} \big] \ket{\alpha\sqrt{R_1}}.
\end{align*}
When we feed this state to the second BS, then the output of the second BS can be calculated by,
\begin{align*}
    \ket{\psi}_2 = & \,  _{b}\bra{1}\hat{U}_{2} \, \ket{\psi}_{1a} \ket{1}_b, \\ 
    = & \, \hbox{N}_2 \, e^{-\abs{\alpha}^2(1-R_1R_2)/{2}} \bigg\{\sqrt{R_1R_2} + \Big[-R_1(1-R_2)
    \\ & -R_2(1-R_1)+(1-R_1)(1-R_2)\Big] \,\alpha \,\hat{a}^{\dagger} 
    \\ & + \sqrt{R_1R_2}(1-R_1)(1-R_2) \, \alpha^2 \, \hat{a}^{\dagger2}\bigg\} \ket{\alpha\sqrt{R_1R_2}}.
\end{align*}
This result can be generalized to $l$ beam splitters and noted as,
\begin{align}
\ket{\Psi}_l = & \hbox{N}_l \, \dfrac{e^{-\frac{\abs{\alpha}^2}{2}(1-X_0)}}{\sqrt{X_0}} \sum^{l}_{m=0} \hbox{T}_m (\hat{a}^{\dagger} \alpha \sqrt{X_0}) \nonumber
\\ & \times (-1)^m X_m \ket{\alpha \sqrt{X_0}}
\label{Eq.Gen}
\end{align}
where, $\hbox{T}_n(y) = \sum_{k=0}^{n}\begin{Bmatrix}
n\\ 
k
\end{Bmatrix} y^k $ is Touchard polynomial \cite{Touchard1939, Chrysaphinou1985} and  
$\begin{Bmatrix}
\bullet\\ 
\bullet
\end{Bmatrix}$ is the Stirling number of the second kind or Stirling partition number \cite{Moser1958,Rennie1969}. Further, 
\begin{align*} 
X_k = \hbox{coefficient of } x^k \hbox{ in } \prod_{i=1}^{l} [R_i+ (1-R_i)x].
\end{align*}
%%%%%%%%%%%%%%%%%%%%%%%%%%%%%%%%%%%%%%%%%%%%%%
\begin{figure}[t]
\centering
         \includegraphics[scale=0.38]{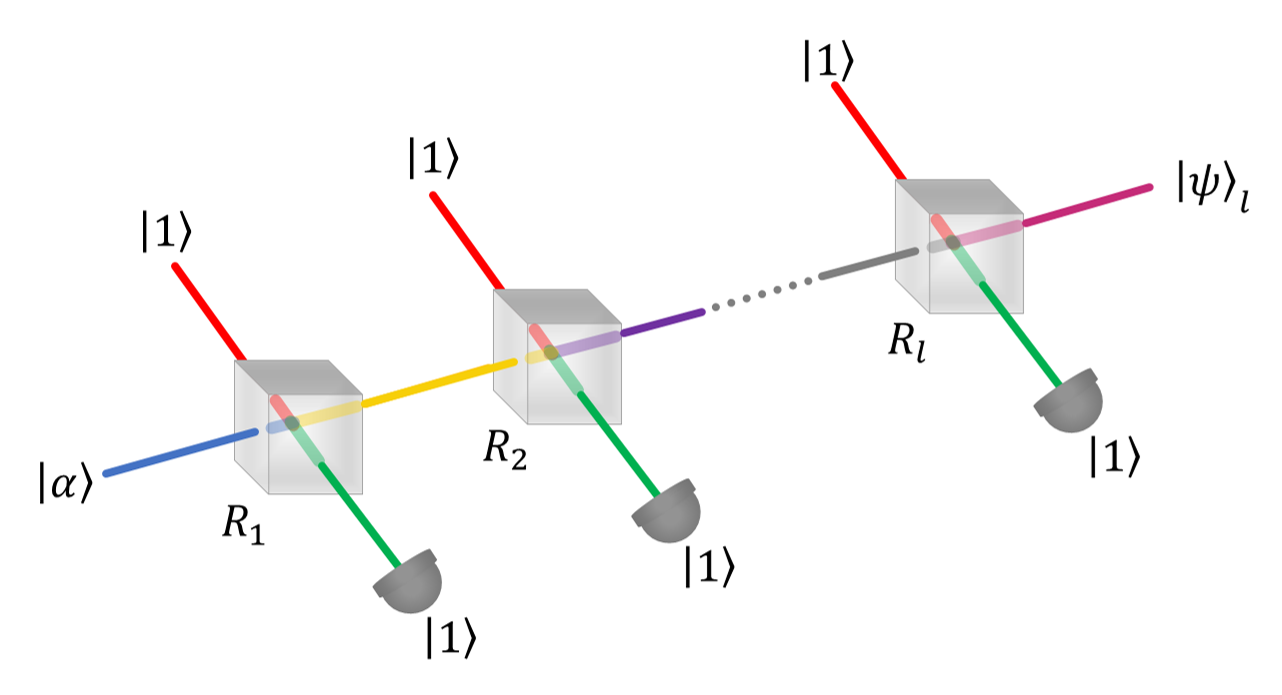}
         \caption{A coherent state and a single photon state are directed into the input ports of $\hbox{BS}_1$. When a single photon is conditionally measured at one of $\hbox{BS}_1$'s output ports, $\hbox{DQ}_1$ is produced at the other port. This $\hbox{DQ}_1$ is subsequently introduced to $\hbox{BS}_2$, where it undergoes a catalysis operation, yielding $\hbox{DQ}_2$ and so on.}
\end{figure}
%%%%%%%%%%%%%%%%%%%%%%%%%%%%%%%%%%%%%%%%%%%%%%%
%%%%%%%%%%%%%%%%%%%%%%%%%%%%%%%%%%%%%%%%%%%%%%%%%%%%%%%%%%%%%%%%%
\begin{figure*}[t]
\begin{center}
%\vspace{-0.5cm}  \hspace{-1.5cm}
\includegraphics[width=17cm,height=6cm]{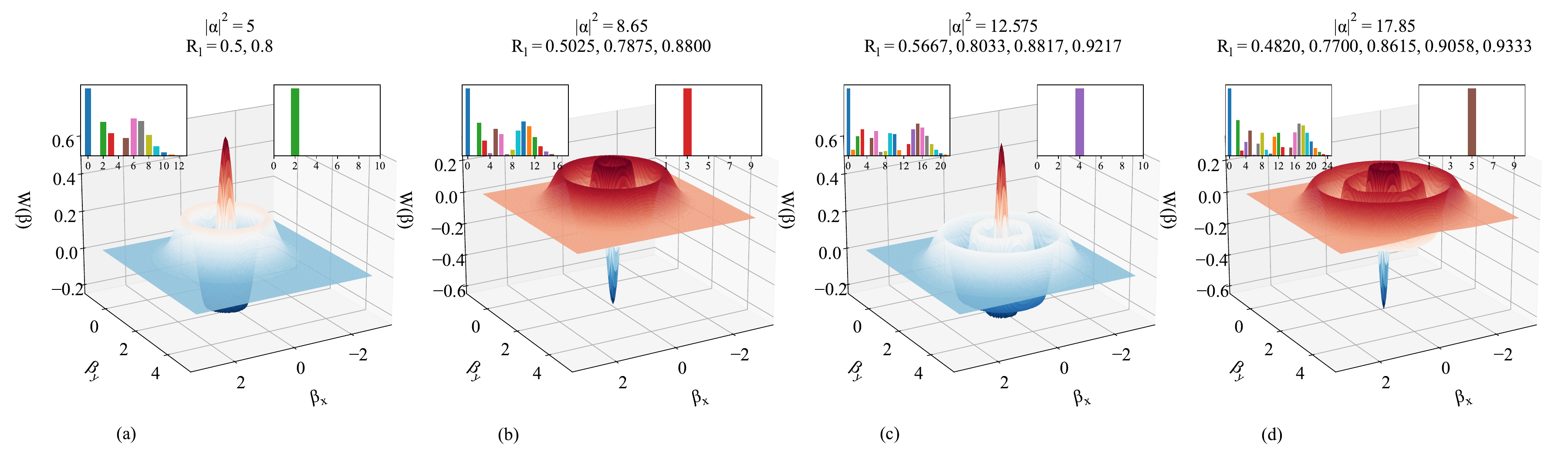}
\caption{The Wigner function of the created photon states with unit fidelity. (a), (b), (c) and (d) correspond to $\ket{2}$, $\ket{3}$, $\ket{4}$ and $\ket{5}$ with displacement respectively.}
\label{Fig.Wig}
\end{center}
\end{figure*}
%%%%%%%%%%%%%%%%%%%%%%%%%%%%%%%%%%%%%%%%%%%%%%%%%%%%%
In Eq. \ref{Eq.Gen}, we can observe the reduction of coherent amplitude due to the measurement. Thus, the input mean photon number is not conserved at the output. Using this equation, it is difficult to characterize the state generation explicitly. Hence, we introduce the displaced qudits form of this output and it is computed as,
\begin{align}
    \ket{\psi}_l = \, \hat{D} (\alpha\sqrt{X_0}) \sum_{p=0}^{l} A_{lp} \ket{p}.
    \label{Eq.CasDQ}
\end{align}
Here, the generation of finite superposed Fock states up to $l$ dimension is clearly visible. Thus, it is also possible to produce the individual Fock states with displacement up to $l-$Fock states. In other words, using $l$ cascaded BS offers the generation $l(=n)$ Fock states. For $l=1,2,3$, the states are represented as displaced qubits $(\hbox{DQ}_{1})$, displaced qutrits $(\hbox{DQ}_{2})$ and displaced ququarts $(\hbox{DQ}_{3})$ respectively. The superposition coefficient $A_{lp}$ is estimated as,
\begin{align}
    A_{lp} = & \hbox{N}_l \,  \sum_{m=p}^{l} (\alpha^*\sqrt{X_0})^{-p} (-1)^m X_{m} \nonumber
    \\ & \times \sum_{k=0}^{m} \begin{Bmatrix}
m\\ 
k
\end{Bmatrix} \binom{k}{p} (\abs{\alpha}^2 X_{0})^k.
\end{align}
The displacement operator, induces sign changes in odd number Fock states due to their intrinsic odd parity and the structure of the operator’s exponential terms. This property is significant in quantum optics for manipulating interference and entanglement in displaced states. The normalization constant is calculated as,
\begin{align}
    \hbox{N}_l = & \bigg \{e^{-\abs{\alpha}^2(1-X_0)} \frac{1}{X_0} \, \sum_{p=0}^{l} \bigg| \sum_{m=p}^{l} (\alpha^*\sqrt{X_0})^{-p} \nonumber
    \\ & \times (-1)^m X_{m} \sum_{k=0}^{m} \begin{Bmatrix}
m\\ 
k
\end{Bmatrix} \binom{k}{p} (\abs{\alpha}^2 X_{0})^k \bigg|^2 \bigg\}^{-\frac{1}{2} \, }.
\end{align}
In this work, we focus on the production of arbitrary quantum states with higher fidelity. The fidelity $(\mathcal{F})$,  is used as measure to quantify the similarity between the DQ's density operator $(\hat{\rho}_{l}=\ket{\psi}_{ll}\bra{\psi})$ and the desired state's density operator $(\hat{\rho}_d)$. The fidelity formula is,
\begin{equation}   
    \mathcal{F} = \hbox{Tr}(\hat{\rho}_{l} \, \hat{\rho}_d).
\end{equation}
Unit fidelity means that the created DQ and desired state match exactly, which are one and same. The success probability (SP) for generating the desired states with our cascaded protocol  can be computed by, 
\begin{equation}
    \hbox{S}_{p} = \abs{N_{l}}^2.
\end{equation}
%%%%%%%%%%%%%%%%%%%%%%%%%%%%%%%%%%%%%%%%%%%%%%%%%%%%%%%%%%%%%%% 
%%%%%%%%%%%%%%%%%%%%%%%%%%%%%%%%%%%%
The Wigner function of DQ is derived as,
\begin{align*}
W(\beta) = & \, e^{2\abs{\beta}^2-\abs{A}^2} \frac{2}{\pi} \sum_{p,q=0}^{l} \frac{A_{lp}A_{lq}^{*}}{\sqrt{p!}\sqrt{q!}} \sum_{u=0}^{p} \binom{p}{u}  (\alpha^*\sqrt{X_0})^{p-u}
\\ & \times \sum_{v=0}^{q} \binom{q}{v} (\alpha \sqrt{X_0})^{q-v} 
\hbox{H}_{u,v} (C^*,C).
\end{align*}
Here, $C=\alpha\sqrt{X_0}-2\beta$. The presence of two variable Hermite polynomial in the Wigner function expresses the non-Gaussian nature of the cascaded output  \cite{Gorska2019}. 

Since the amplitude coefficients $A_{\ell p}$ involve sequential nonlinear equations, finding direct analytical solutions for the PND equivalence of each state are inconceivable. For each cascaded configuration, the amplitude coefficients $A_{\ell p}$ are extracted as functions of the system parameters. Specifically, for a given level $\ell$, the coefficients correspond to all photon number indices $p = 0, 1, 2, \ldots, \ell$. The coherent amplitude parameter $\alpha$ is varied over the continuous range $0 \leq \alpha \leq 5$, while each reflectivity parameter $R_i$ is independently scanned within the interval $0 \leq R_i \leq 1$.

This scanning procedure yields a complete $(\ell + 1)$-dimensional dataset of amplitude coefficients associated with each cascaded configuration. 
For instance, in a system with three cascaded stages ($\ell = 3$), the amplitudes
\[
|A_{30}|^2, \quad |A_{31}|^2, \quad |A_{32}|^2, \quad |A_{33}|^2
\]
are computed for every combination of parameters within the four dimensional space defined by $\{\alpha, R_1, R_2, R_3\}$. 
Each computed dataset is systematically recorded and stored for subsequent optimization and state construction. In this article, all nonclassical states are generated exclusively by employing the aforementioned methodology. 
%%%%%%%%%%%%%%%%%%%%%%%%%%%%%%%%%%%%%%%%%%%%%%%%%%%%%%%%%%%%%%%%%%%%
\section{Fock states}
\label{sect.Fock}
Fock states also known as photon number states can be characterized by a well defined photon number \cite{Loudon2000, Gerry2004}. The negative values in their Wigner functions evidence the intrinsic non-Gaussian character \cite{Genoni2007, Walschaers2021}. Higher photon states are need for the applications of linear optical quantum computing \cite{Dowling2003, Kok2007}, quantum simulations \cite{Sturges2021} and quantum metrology \cite{Deng2024}. Although single photon states can be produced by heralding spontaneous parametric down conversion \cite{Hong1986, Burnham1970} and spontaneous four wave mixing \cite{Soller2011}, the realization of higher Fock states is hindered by their low probability of occurrence. Higher Fock states are realized by performing conditional measurement on the nonlinear process \cite{Ourjoumtsev2006b, Cooper2013}. 
%%%%%%%%%%%%%%%%%%%%%%%%%%%%%%%%%%%%%%%%%%%%%
\begin{figure*}[t]
\begin{center}
%\vspace{-0.5cm}  \hspace{-1.5cm}
\includegraphics[width=17cm,height=13cm]{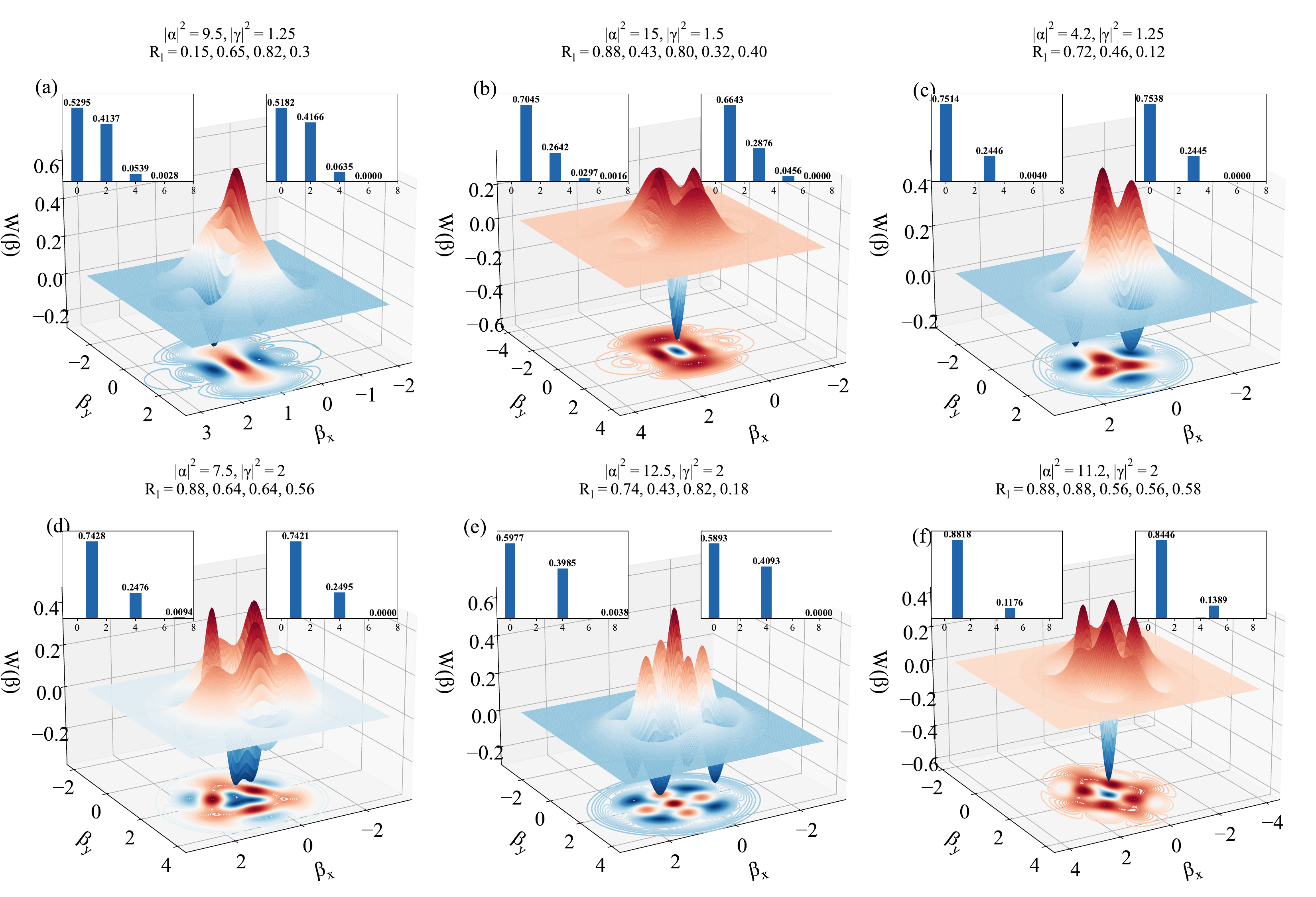}
\caption{The Wigner functions for the generated states are depicted as follows: (a) even cat states with $h=0$, (b) odd cat states with $h=1$. For three headed cat states, the Wigner functions are shown in (c) for $h=0$ and (d) for $h=1$. Similarly, the Wigner functions of compass states are illustrated in (e) for $h=0$ and (f) for $h=1$.}
\label{Fig.LSCS}
\end{center}
\end{figure*}
%%%%%%%%%%%%%%%%%%%%%%%%%%%%%%%%%%%%%%%%%%%%%%%%%%%%%
Our present study explores the generation of higher Fock states with displacement using a similar cascaded BS arrangements \cite{Escher2005}, except with modified input scheme that combines coherent states and single photon. By implementing single-photon catalysis at each beam splitter stage and optimizing the input parameters, we aim to produce displaced Fock states with unit fidelity. Surprisingly, our chosen setup also enables the generation of higher Fock states but with some displacement. The displacement operator $\hat{D}(\alpha\sqrt{R_l})$ is a Gaussian operation, so it does not affect the non-Gaussianity generated in the Fock states.

First, let us describe the methodology of finding the creation of higher photon number states in the respective cascaded setup. We are trying to produce $n$ photon states with $l$ BSs. In other words, if $l=2$, then we are aiming to create $\ket{2}$ successively, if $l=5$ then creating the $\ket{5}$ with some displacement. First, the generation of $\hat{D}(\alpha \sqrt{R_1R_2})\ket{2}$ using $l=2$ cascaded BSs with single photon catalysis on each and all. From Eq. \ref{Eq.CasDQ}, we can note that if we make $\abs{A_{22}}^2 = 1$ then it results in $\ket{2}$. By applying this condition, the obtained expression is,
\begin{align*}
    \abs{A_{22}}^2 = \abs{\hbox{N}_2}^2 \alpha^4 R_1^2R_2^2(1-R_1)^2(1-R_2)^2 = 1.
\end{align*}
It is certain for the values of $\abs{\alpha}^2=5, $ and reflectivities of $0.5$ and $0.8.$ Thus $\hat{D}(\sqrt{2})\ket{2}$ is generated. The success probability for the creation this $\ket{2}$ state is $1\%$, see tab. \ref{tab:FSQLSCS}. To generate the state $\ket{3}$, we must use three cascaded beam splitters. It is known that to produce $\ket{3}$, we first need to generate $\ket{2}$ at the output of the second beam splitter. This step is essential, as it allows us to combine the nonclassicality of an additional $\ket{1}$ with $\ket{2}$ to create $\ket{3}$. Therefore, we can leverage prior knowledge of the reflectivity values required for the second beam splitter to synthesize $\ket{2}$. Specifically, the reflectivities of the first two beam splitters in the three beam splitter setup should be close to those optimized for $\ket{2}$ generation. This insight enables us to estimate the reflectivity values for the entire setup. It was found that with reflectivity values of $0.5025, 0.7875,$ and $0.8800$, we successfully generated $\hat{D}(\sqrt{3})\ket{3}$.

The generation of other Fock states was computed similarly. By enforcing the condition $\abs{A_{ll}}^2=1$, we can determine the input parameters. The creation of higher photon states with unit fidelity is summarized in Table \ref{tab:FSQLSCS}, which lists the tuned coherent amplitudes and reflectivity values. All states are produced with their respective displacement operations, denoted as $\hat{D}(\alpha \sqrt{X_0})$. Generally, the displacement for each state is approximately $\sqrt{n}$, i.e., $\hat{D}(\sqrt{n})$, for the corresponding Fock state $n$. 

The Wigner functions of the created displaced $\hat{D}(\alpha\sqrt{X_0})$ Fock states are plotted in the fig. \ref{Fig.Wig}: (a) $\ket{2}$, (b) $\ket{3}$, (c) $\ket{4}$ and (d) $\ket{5}$. The respective coherent parameter $\alpha$ and BS reflectivies $R_l$ accounted in the generation of Fock states are mentioned in the figure. Each main panel shows the generation of these states with unit fidelity. The left insets in each sub-figure present the PND of the displaced state, highlighting how the displacement modifies the photon number statistics. The right insets depict the PND of the corresponding Fock state without displacement, which is the original photon number distribution from the cascaded setup. 
%}
%%%%%%%%%%%%%%%%%%%%%%%%%%%%%%%%%%%%%%%%%%%%%%%%%%%%%%%%%%%%%%% 
%%%%%%%%%%%%%%%%%%%%%%%%%%%%%%%%%%%%%%%%%%%%%%%%%%%%%%%%%%%%%
\section{Family of Schrodinger cat states}
\label{sect.FSCS}
%%%%%%%%%%%%%%%%%%%%%%%%%%%%%%%%%%%%%%%%%%%%%%%%%%%%%%%%%%%%%%%%%
The linear superposition of coherent states (LSCS), also known as a family of Schrodinger cat states, is a macroscopic superposition of quantum states. This family comprises Schrodinger cat states, three headed cat states, compass states and other multi component cat states. These states are playing a central role in quantum error correction such as bosonic error correcting codes \cite{Cochrane1999,Michael2016} and precision metrology \cite{Mirrahimi2014,Tan2024}. LSCS can be generated through various processes like using photon subtraction from squeezed states \cite{NeergaardNielsen2006}, conditional measurements on entangled fields \cite{Dakna1997}, and deterministic schemes \cite{Hastrup2019,Thekkadath2020}. Specifically, Compass states exhibit sub-Planck scale interference patterns in phase space, which enhances the sensitivity of measurement \cite{Zurek2001,Tan2024}. LSCS is expressed by,
\begin{align}
    \ket{\phi}_{g,h} = N_{g,h} \sum_{r=0}^{g-1} e^{-i2\pi rh/g} \ket{\gamma e^{i2\pi r/g}}.
\end{align}
Here, $\gamma$ refers the coherent amplitude of this system and $g$ limits the number of superposition terms in the system. For example, at $g=1$, the state expresses the coherent states. $h$ denotes the phase angle of $(2\pi r/g)$ between the superposition terms it includes the values upto $h=0,1,2,...g-1$. Normalization constant $N_{g,h}$ is computed as,
\begin{align*}
    N_{g,h} = \frac{1}{\sqrt{g}} \abs{\sum_{r=0}^{g-1} e^{-i2\pi rh/g} e^{-\abs{\alpha}^2(1-e^{i2\pi r/g})}}^{\frac{1}{2}}.
\end{align*}
In this work, we adopt the following methodology. First, we compute the PNDs of the respective LSCS for the specified values of $\gamma$. Subsequently, we match the relevant amplitudes to $\abs{A_{lp}}^2$ and thereby determine the input parameters of the cascaded protocol, such as $\alpha$ and $R_l$. 
\subsection{Schrodinger Cat states}
Along $g=2$, the states comprises the even and odd cat states at $h=0$ and $1$ respectively. The even cat is derived as, $\ket{\phi}_{2,0} = N_{2,0} \big[\ket{\gamma}+\ket{-\gamma}\big]$. Fig. \ref{Fig.LSCS}(a) interprets even cat state of amplitude $\abs{\gamma}^2=1.25$ creation using four cascaded setup upto $98\%$ fidelity. The odd cat is derived as, $\ket{\phi}_{2,1} = N_{2,1} \big[\ket{\gamma}-\ket{-\gamma}\big]$. Odd cat state of amplitude $\abs{\gamma}^2=1.5$ production through five cascaded BS is pictured, see Fig. \ref{Fig.LSCS}(b).
\subsection{Three headed cat states}
Three headed cat states are the superposition of three coherent states with different phase angles. When $g=3$ in LSCS, the state produces the three headed cat states. For the phase angles of $h=0$, $\ket{\phi}_{3,0} = N_{3,0} \big[ \ket{\gamma}+\ket{\gamma e^{i2\pi/3}}+\ket{\gamma e^{i4\pi/3}}\big]$. Employing a configuration of three cascaded stages, we generated three headed cat states with an amplitude of $1.25$, as shown in Fig.~\ref{Fig.LSCS}(c). At $h=1$, the another phase rotated three headed cat state, $\ket{\phi}_{3,1} = N_{3,1} \big[ \ket{\gamma}+e^{-i2\pi/3}\ket{\gamma e^{i2\pi/3}}+e^{-i4\pi/3}\ket{\gamma e^{i4\pi/3}}\big]$ is expressed. Fig. \ref{Fig.LSCS}(d) interprets this state $\ket{\phi}_{3,1}$ creation for the amplitude of $1.75$ by implementing four cascaded setup.  
\subsection{Compass states}
Compass states are the four component superposition of coherent states, which can also be viewed as superpositions of two cat states. For $g=4$, and $h=0,1,2,3$, the states exhibits the compass state with different optical phases. At $g=4$ and $h=0$, the acquired compass state is, $\ket{\phi}_{4,0} = N_{4,0} \big[\ket{\gamma}+\ket{-\gamma}+\ket{i\gamma}+\ket{-i\gamma}\big]$. The compass state generations using our setup is assured. Fig. \ref{Fig.LSCS}(e) depict the Compass of amplitude $\abs{\gamma}^2=2$ and the phase rotation of $h=0$ for around $98\%$ fidelity. The four blue valley regions which portray the four directions (north, east, south, west) i.e. compass natures as mentioned in the article \cite{Zurek2001}. Another compass state for the same amplitude with different phase rotation $h=1$, $\ket{\phi}_{4,1} = N_{4,1} \big[\ket{\gamma}-\ket{-\gamma}-i\ket{i\gamma}+\ket{-i\gamma}\big]$ is created by using five cascaded setup is visualized in the figure. \ref{Fig.LSCS}(f). 
%%%%%%%%%%%%%%%%%%%%%%%%%%%%%%%%%%%%%%%%%%%%%%%%%%%%%%%%%%%%%%
%%%%%%%%%%%%%%%%%%%%%%%%%%%%%%%%%%%%%%%%%%%%%%%%%%%%%%%%%%%%%%
\section{Finite Superposition of Number States}
\label{sect.FSNS}
One primary method using only linear optics to generate finite superposition of number states (FSNS) is the quantum scissors technique which truncates quantum states using beam splitters and photon detection \cite{Pegg1998, Barnett1999}. However, its ability to create FSNS is limited up to certain states, it is unfeasible for generating arbitrary FSNS. Despite that arbitrary FSNS are essential for synthesizing the squeezed states and various nonclassical states.  

The generation of superposed states using a cascaded arrangement of BS with coherent state and single photon as inputs and by changing the photon counting in the detector \cite{Miller2019}. In this section, we try to get those states even by fixing the detectors to count single photons only. Here, we followed the same methodology used in the previous section, the difference is we try to generate $n-$ superposition states with $l-$ cascaded beam splitters. For example, if we need to create a $3-$ finite dimensional states $A_{0}\ket{0}+A_{2}\ket{2}+A_{3}\ket{3}$, then we have to choose the $3-$ cascaded arrangement of beam splitters. The FSNS is taken as, 
$\ket{\chi}_n = \sum_{p=0}^{n} B_{np} \ket{p}$. 
%%%%%%%%%%%%%%%%%%%%%%%%%%%%%%%%%%%%%%%%%%%%%%%%%%%%%%%%%%

\subsection{Optimal squeezed states in FSNS}
To enhance phase precision in weak signal detection, nonclassical states rely on squeezing features, which minimize the uncertainty in one observable at the expense of another \cite{Loudon1987, Walls1983, Caves2013, Schnabel2017}. Evidently, FSNS, which combine a finite set of photon number states, exhibit inherent squeezing properties \cite{Wodkiewicz1987, Orllowski1991, Figurny1993}. Initially, the Jaynes-Cummings model was used to create FSNS \cite{Lee1993, Wodkiewicz1987}, but these often lacked ideal squeezing. Recently, another approach using conditional measurement on a coherent state and a photon number has been employed to produce FSNS \cite{Esakkimuthu2025}. This method achieves optimal squeezing for superpositions with up to two photon number states by utilizing linear optics solely.

The present setup also produces finite Fock state superposition even though all the input states are unsqueezed states. To calibrate the optimal squeezing generation, we compute the quadrature squeezing properties of DQ. The quadrature components of the optical field are defined as $\hat{X}= \frac{(\hat{a}+\hat{a}^{\dagger})}{\sqrt{2}}$ and $\hat{P}= \frac{(\hat{a}- \hat{a}^{\dagger})}{i\sqrt{2}}$. The squeezing properties of the states is identified when any of the quadrature variances goes the less than $1/2$ by complementing the other quadrature variance to satisfy the condition $\expval{\Delta\hat{X}^2}\expval{\Delta\hat{P}^2} = 1/4$.
In our case, the minimum quadrature variance of DQ occurs at $\hat{X}$ quadrature. The higher order moments of DQ is evaluated as,
\begin{align} 
    \expval{\hat{a}^{\dagger l}\hat{a}^{s}} = & \,  \sum_{p=0}^{l} A_{lp} \sum_{u=0}^{t} \binom{t}{u} \sum_{v=0}^s \binom{s}{v} A_{l(p-v+u)}^{*} \nonumber
    \\ & \times (\alpha^* \sqrt{R})^{l+s-u-v} \, \frac{[p!(p-v+u)!]^{1/2}}{(p-v)!}.
    \label{Eq.Mom}
\end{align}
Here, if the index of the coefficients possesses a negative value, it is considered as zero i.e., $(A_{lp(-ve)} = 0)$. The expressions of first and second-order moments can be obtained from Eq. \ref{Eq.Mom}, by substituting $l=0, s=1$ and $l=0, s=2$ respectively. 

Before examining the squeezing properties of $\hbox{DQ}$, we note two points: (i) the displacement operator $\hat{D}(\alpha\sqrt{R_l})$ does not affect squeezing in the constructed superposed states \cite{Lee1989}, as it cannot generate nonclassicality; (ii) optimal squeezing depends on the choice of superposition coefficients. For Fock state superpositions $\ket{\chi}_n = \sum_{p=0}^{n} B_{np} \ket{p}$ \cite{Orlowski1995}, the maximum squeezing for $\ket{\chi}_1 = B_{10}\ket{0}+B_{11}\ket{1}$ is $0.3750$ when $\abs{B_{10}}^2=3/4$. Similarly, for $\ket{\chi}_2 = B_{20} \ket{0}+ B_{21}\ket{1}+B_{22}\ket{2}$it is $0.2753$ with $B_{20} = 0.9530, B_{21} = 0$ and $B_{22} = - 0.3030$. Within this, we try to produce the finite superposed states which have the equivalent maximum squeezing within the corresponding $l-$cascaded parallels to $n-$ superposed states. 
%%%%%%%%%%%%%%%%%%%%%%%%%%%%%%%%%%%%%%%%%%%%%%%%%%%%%%%%%%%%%%%%%%%%%%%%%%%%%%%%%%%%%%%%
\begin{table}[htbp]
\centering
\caption{Minimum $(\Delta x)^2$ obtained in DQ.}
\begin{adjustbox}{width=1\columnwidth}
\begin{tabular}{c c c c c c}
\hline
$l$ & min$(\Delta x)^2$ & (dB) & $\abs{\alpha}^2$ & $R_l$ & SP \\
\hline
\multirow{2}{*}{1} 
    & \multirow{2}{*}{0.3750} 
    & \multirow{2}{*}{1.25} 
    & $\dfrac{1}{4R}\left[\sqrt{3} \pm \sqrt{\dfrac{3+R}{1-R}} \right]^2$ 
    & \multirow{2}{*}{$-$} 
    & \multirow{2}{*}{0.3388} \\
    & & & & & \\
2 & 0.2753 & 2.59 & 12.4 & 0.92, 0.49 & 0.0022 \\
3 & 0.2297 & 3.38 & 14.6 & 0.83, 0.95, 0.78 & 0.0015 \\
4 & 0.1902 & 4.20 & 6.55 & 0.80, 0.36, 0.65, 0.12 & $5.03\!\times\!10^{-5}$ \\
5 & 0.1666 & 4.77 & 11.0 & 0.62, 0.25, 0.79, 0.86, 0.27 & $6.82\!\times\!10^{-7}$ \\
\hline
\end{tabular}
\end{adjustbox}
\label{tab:SQ}
\end{table}
%%%%%%%%%%%%%%%%%%%%%%%%%%%%%%%%%%%%%%%%%%%%%%%%%%%%%%%%%%%%%

Minimum quadrature variance $(\Delta x)^2$ achieved in DQ for various $n-$FSNS. The table lists the minimum value of $(\Delta x)^2$ in both absolute units and decibels (dB) in Table. \ref{tab:SQ}. The input coherent amplitudes and the set of $R_l$ parameters is also noted in the table. For $n=1$, $(\Delta x)^2_{\text{min}} = 0.3750$ (1.25 dB) is obtained when $|\alpha|^2$ satisfies the given analytic expression, with $R_l = 3/|\alpha|^2$ and $SP = 0.3388$. For $n=2$--$5$, progressively lower $(\Delta x)^2_{\text{min}}$ values and higher squeezing levels are observed (up to 4.77 dB for $n=5$), with corresponding $|\alpha|^2$ and $R_l$ values giving optimal squeezing. The SP decreases notably with increasing $n$, indicating reduced squeezing probability at higher photon numbers. In the previous works conditional measurements with coherent and arbitrary photon states, the creation of higher squeezing is not possible beyond $2-$ superposed states \cite{Esakkimuthu2025}. In this case, the method allows for the possibility of approaching the maximum achievable squeezing in the given FSNS. This is owing to the increase in tuning parameters in the setup i.e. more reflectivities $R_l$, which makes it possible to generate the desired arbitrary quantum states.
%%%%%%%%%%%%%%s%%%%%%%%%%%%%%%%%%%%%%%%%%%%%%%%%%%%%%%%%%
\begin{figure*}[t]
    \begin{subfigure}{0.325\textwidth}
         \includegraphics[scale=0.16]{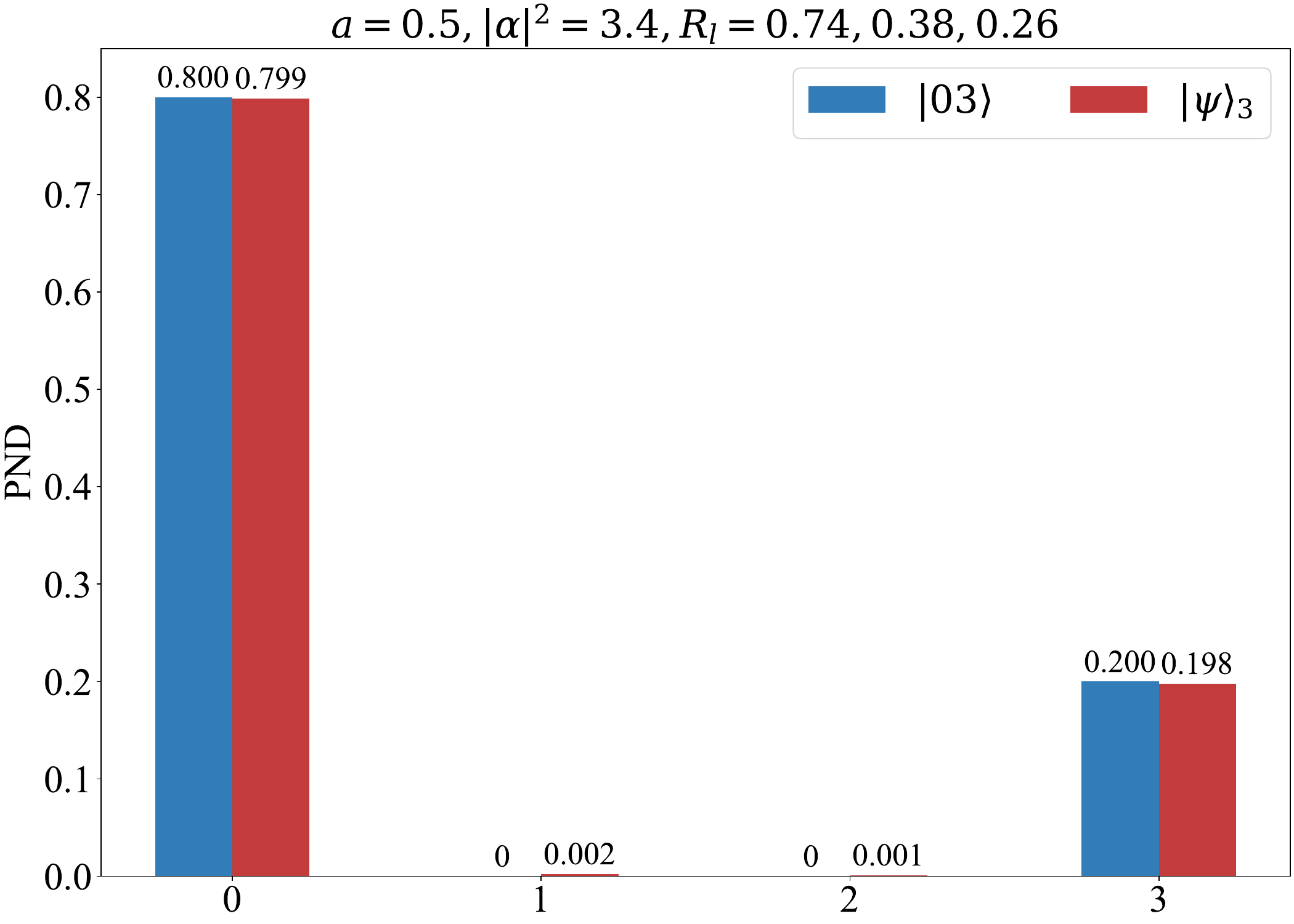}
         \subcaption[]{}
         \label{Fig.N03bar}
     \end{subfigure}
     \hfill
     \begin{subfigure}{0.325\textwidth}
         \includegraphics[scale=0.16]{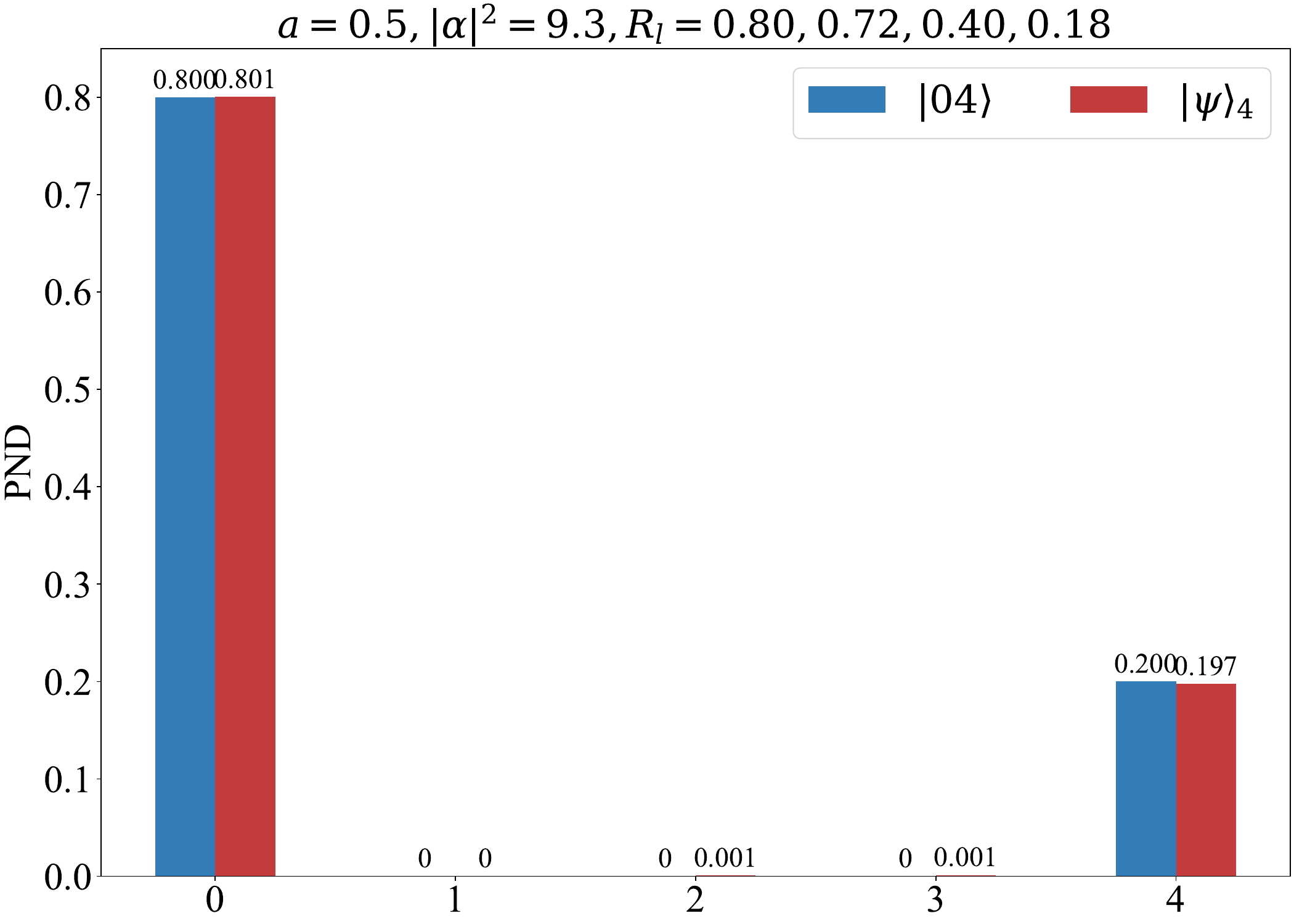}
         \subcaption[]{}
         \label{Fig.N04bar}
     \end{subfigure}
     \hfill
     \begin{subfigure}{0.325\textwidth}
         \includegraphics[scale=0.16]{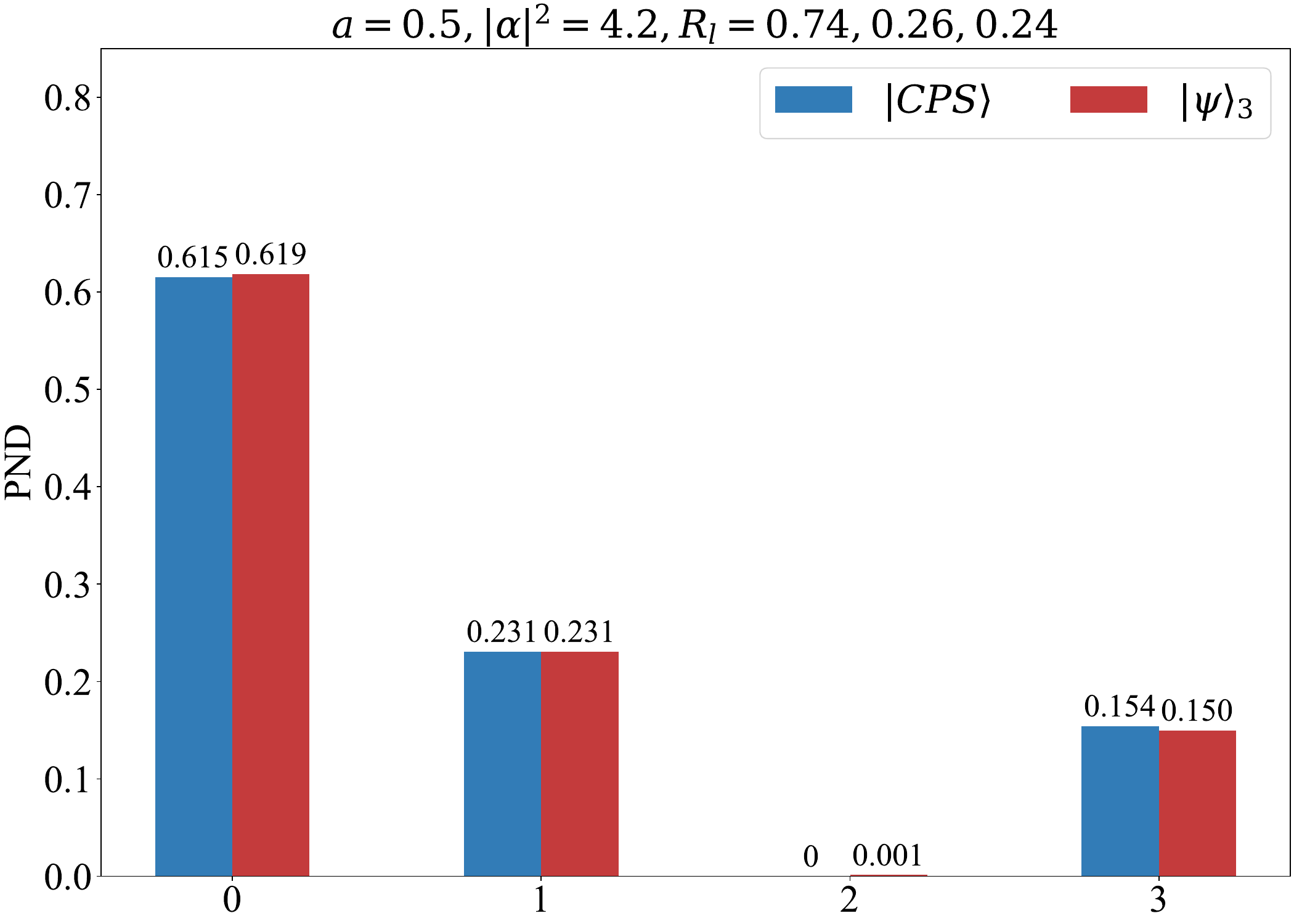}
         \subcaption[]{}
         \label{Fig.CPSbar}
     \end{subfigure}
     \hfill
     \caption{The bar plots portray the comparison of photon probability amplitudes PND between the target states and the cascaded output states. PND of the states between (a) $\ket{03}$ and $\ket{\psi}_3$, (b) $\ket{04}$ and $\ket{\psi}_4$ and (c) $\ket{CPS}$ and $\ket{\psi}_3$ respectively.}
     \label{Fig.ONpnd}
\end{figure*}
%%%%%%%%%%%%%%%%%%%%%%%%%%%%%%%%%%%%%%%%%%%%%%%%%%%%%%%%%%%%%%%%%%%
\subsection{GKP states resources}
Finally, we address the arbitrary generation of single mode quantum states using a cascaded protocol. As a concrete example, we focus on the generation of ON resource states and cubic phase states. ON states have been investigated as fundamental resource units for universal quantum computation, serve as a foundational building block for the preparation of Gottesman–Kitaev–Preskill (GKP) states \cite{Sabapathy2018}. These ON states are defined as quantum superpositions of the vacuum state and the $n$-th Fock state,
\begin{align}
    \ket{\hbox{ON}} = \dfrac{1}{\sqrt{1+\abs{a}^2}} (\ket{0}+a\ket{N}).
\end{align}
The Fock basis expansion of a weak cubic phase state is represented as \cite{Sabapathy2019CPS},
\begin{align}
    \ket{CPS} = \dfrac{1}{\sqrt{1+5\abs{a}^2/2}} \Bigg[\ket{0}+ia\sqrt{\frac{3}{2}}\ket{1}+ia\ket{3} \Bigg].
\end{align}
Fig. \ref{Fig.ONpnd} displays the PND correspondence between the target states $\ket{ON}$, $\ket{CPS}$, and $\ket{\psi}_l$. For the specific case $a = 0.5$, the procedure for state generation is shown. Each state is produced with a fidelity of up to $99\%$. The generation of additional states for different values of $a$ is also feasible within our experimental configuration. Fig. \ref{Fig.ONpnd} (a), (b) and (c) portray the PND's of $\ket{03}$, $\ket{04}$ and $\ket{CPS}$ along with the PND amplitudes of the matched cascaded states. 
%%%%%%%%%%%%%%%%%%%%%%%%%%%%%%%%%%%%%%%%%%%%%%%%%%%%%%%%%%%%%%%%
\section{Arbitrary single Mode State Generation}
\label{sec:arbitrary}

The generation of arbitrary single mode nonclassical states is a fundamental task in quantum optics and CV quantum information \cite{Dakna1999b, Fiurasek2005}. The proposed protocol involves the coherent manipulation of optical amplitudes at each cascade level to synthesize a desired superposition of Fock states. The methodology presented below details both the computational framework for generating the amplitude coefficients and the optimization protocol used to reproduce the target photon number distribution. 
%%%%%%%%%%%%%%%%%%%%%%%%%%%%%%%%%%%%%%%%%%%%%%%%%%
\begin{algorithm}[H]
\caption{Generalized $\ell$-Level Amplitude Optimization Protocol}
\label{alg:ampopt}
\begin{algorithmic}[1]
\Require Desired amplitude distribution $\{|A_{\ell p}|^2_{\text{target}}\}$, level $\ell$, and coherent amplitude parameter $\alpha$.
\Ensure Optimal reflectivities $\{R_l\}$ and $\alpha$ satisfying $\sum_p |A_{\ell p}|^2 = 1$.

\Statex \textbf{Step 1: Initialize Parameters}
\State Define symbolic coherent amplitude $\alpha$ and reflectivities $R_1, \dots, R_\ell$.
%\State Compute symmetric polynomials $e_s(R_1, \dots, R_\ell)$.

\Statex \textbf{Step 2: Construct Amplitudes}
\For{$p = 0$ \textbf{to} $\ell$}
  \State Compute $A_{\ell p}$ via the symbolic series definition.
\EndFor

\Statex \textbf{Step 3: Normalize Amplitudes}
\State $N_\ell = \left(\sum_{p=0}^{\ell} |A_{\ell p}|^2\right)^{-1/2}$.
\State Normalize each amplitude: $A_{\ell p} \gets N_\ell \, A_{\ell p}$.

\Statex \textbf{Step 4: Define Objective Function}
\State Define minimization error: 
$$\mathcal{E}(\mathbf{R}, \alpha) = \sum_{p=0}^{\ell} (|A_{\ell p}|^2 - |A_{\ell p}|^2_{\text{target}})^2.$$

\Statex \textbf{Step 5: Optimization}
\State Parameter bounds: $0 \le R_l \le 1$, $0 \le \alpha \le 5$.
\State Solve $\min_{\{R_i\}, \alpha} \mathcal{E}(\mathbf{R}, \alpha)$
\Statex \hspace{1em} subject to $\sum_{p=0}^{\ell} |A_{\ell p}|^2 = 1$.
\Statex \hspace{1em} • Local: Sequential Quadratic Programming using using \texttt{fmincon}.  
\Statex \hspace{1em} • Global: MultiStart or Particle Swarm Optimization.

\Statex \textbf{Step 6: Output Results}
\State Return $\{R_l, \alpha\}$, minimized $\mathcal{E}$, and $A_{\ell p}$.
%\State Plot optimized vs. target amplitudes.

\end{algorithmic}
\end{algorithm}
%%%%%%%%%%%%%%%%%%%%%%%%%%%%%%%%%%%%%%%%%%%%%%%%%%%%%%%%%%%
To synthesize a desired nonclassical state, the PND of the target state is first determined. The objective of the optimization procedure is to identify the input parameters of the cascaded setup namely, the coherent amplitude $\alpha$ and reflectivities $\{R_l\}$—that reproduce an amplitude distribution closely matching the target PND. This goal is achieved through a generalized amplitude optimization protocol, in which the difference between the target and simulated amplitude distributions is minimized under normalization constraints. The detailed computational steps of this optimization framework are described in Algorithm~\ref{alg:ampopt}.
%%%%%%%%%%%%%%%%%%%%%%%%%%%%%%%%%%%%%%%%%%%%%%%%%%
\section{Realistic preparation}
\label{sect.RQOC}
%%%%%%%%%%%%%%%%%%%%%%%%%%%%%%%%%%%%%%%%%%%%%%%%%%%%%%%%%
\begin{figure*}[t]
    \begin{subfigure}{0.325\textwidth}
         \includegraphics[scale=0.17]{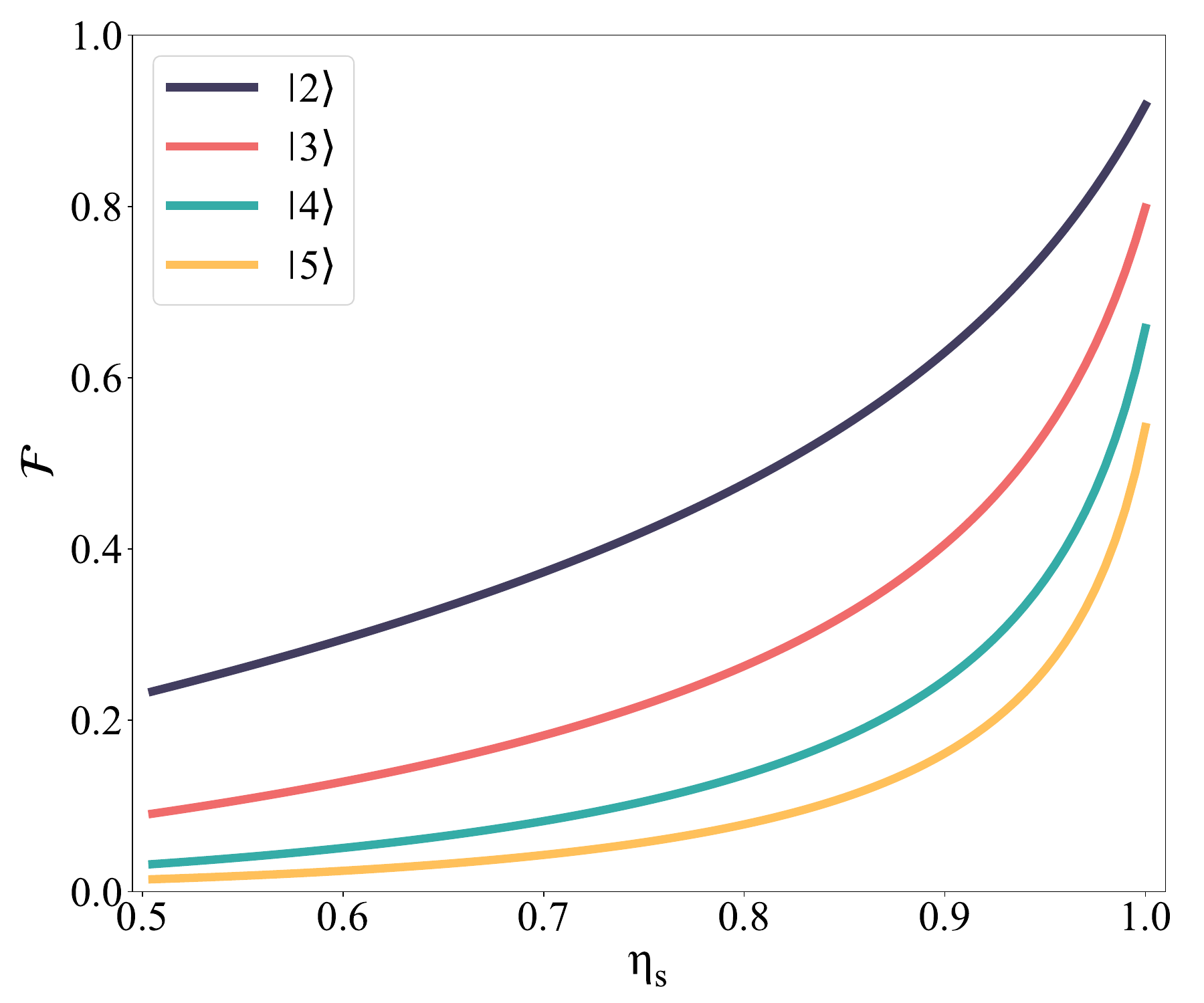}
         \subcaption[]{}
         \label{Fig.RQOCFock}
     \end{subfigure}
     \hfill
     \begin{subfigure}{0.325\textwidth}
         \includegraphics[scale=0.17]{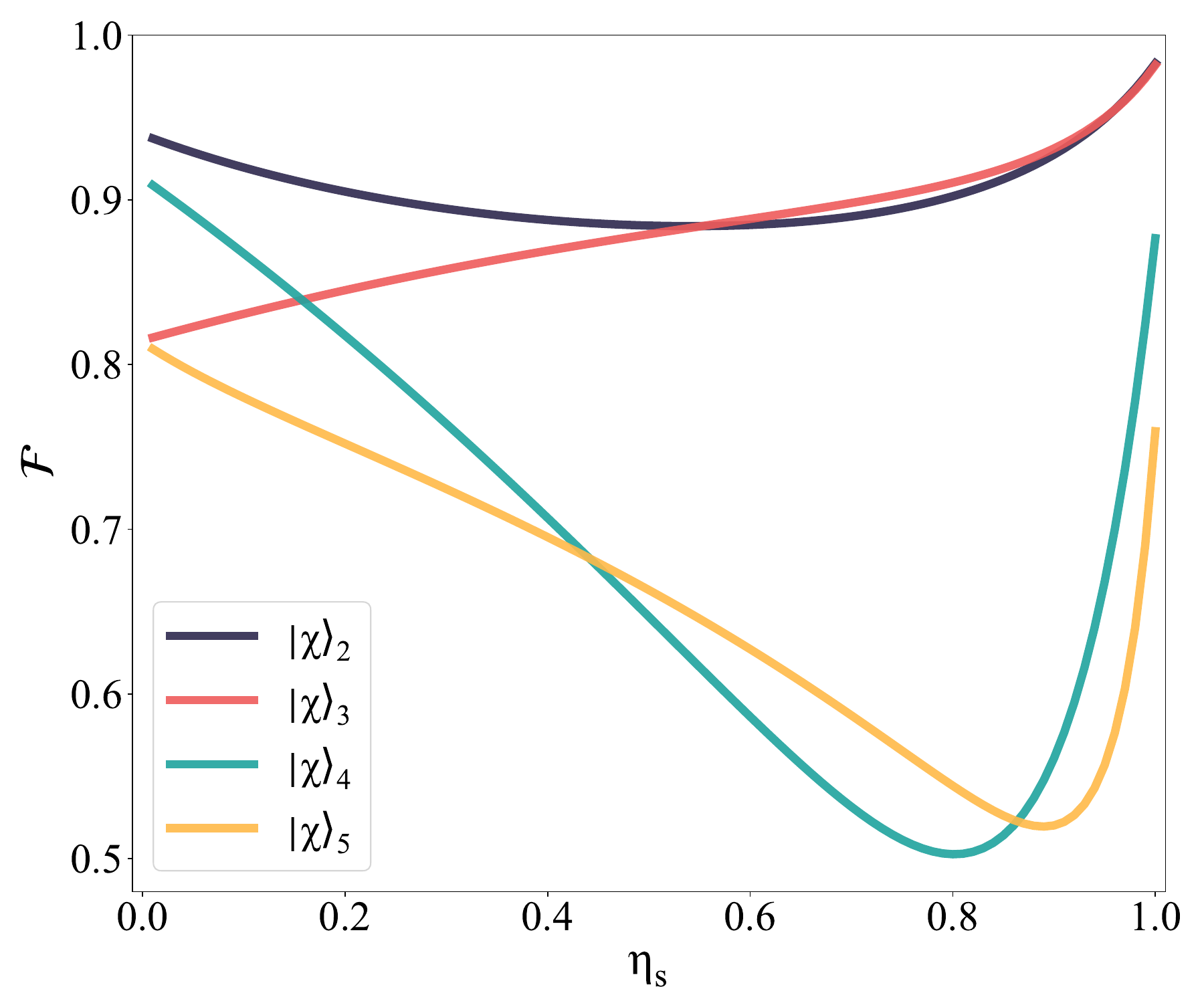}
         \subcaption[]{}
         \label{Fig.RQOCSQ}
     \end{subfigure}
     \hfill
     \begin{subfigure}{0.325\textwidth}
         \includegraphics[scale=0.17]{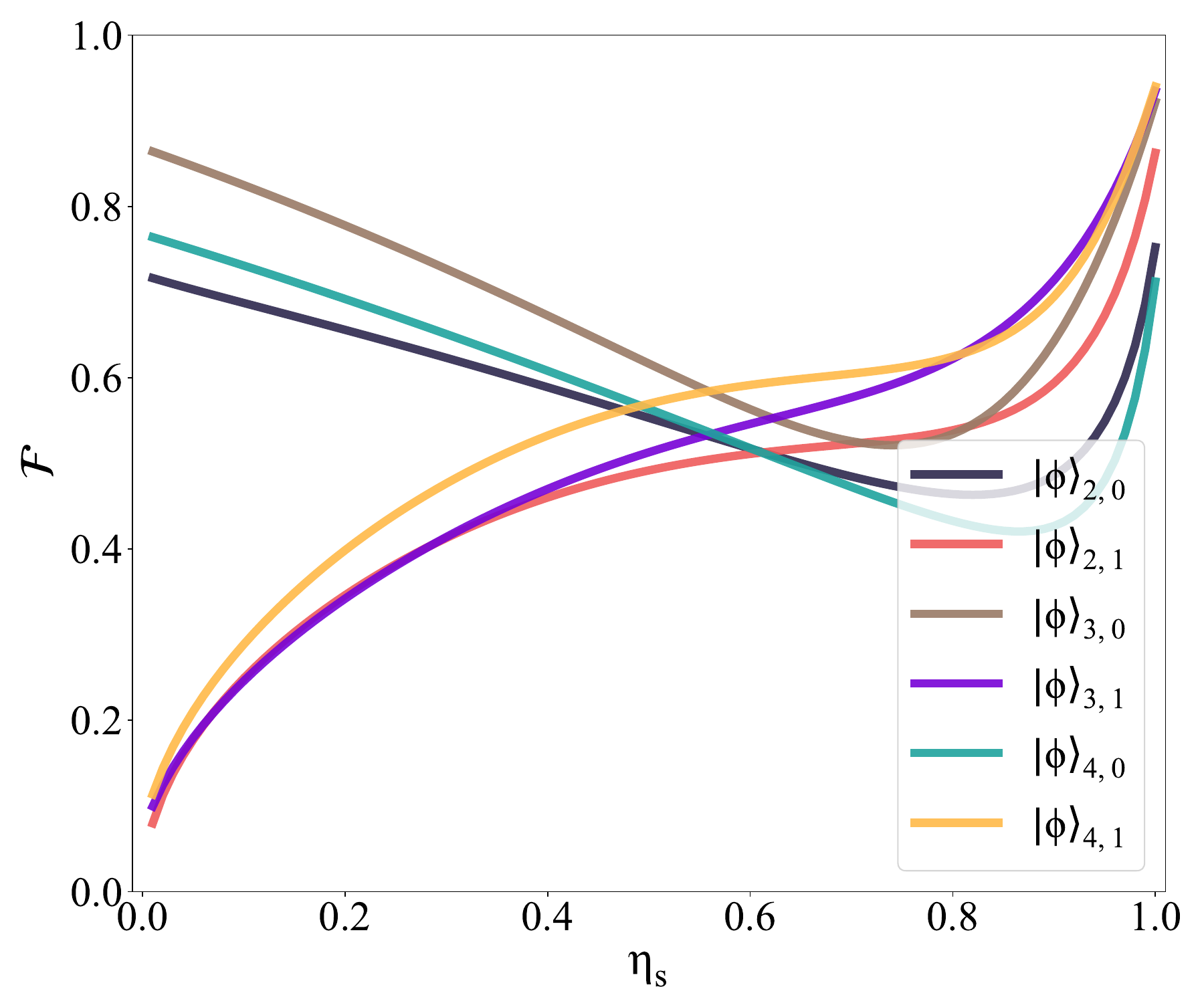}
         \subcaption[]{}
         \label{Fig.RQOCLSCS}
     \end{subfigure}
     \hfill
     \caption{The figures portray the fidelity between the realistic generation and the ideal generation of the created states. Here, detector inefficiency $\eta_d = 0.98$ is fixed and fidelity is taken for various non-unit source efficiency $\eta_s$. (a), (b) and (c) correspond to higher Fock states, optimal squeezed states and LSCS generation respectively.}
     \label{Fig.RQOC}
\end{figure*}
%%%%%%%%%%%%%%%%%%%%%%%%%%%%%%%%%%%%%%%%%%%%%%%%%%%%%%%%%%%%%%%%%%%
In this section, the experimental imperfections in the cascaded protocol for the generation of nonclassical states are taken into account. Two major imperfections are considered: the practical impact of photon detector inefficiency and the mixed state nature of photon sources. Neglecting dark counts, the POVM for a photon number resolving detector identifying a single photon state is given by \cite{Scully1969},
\begin{equation}
    \hat{\pi}_1 = \sum_{k=1}^{\infty}\, {k} \,  \eta_d (1-\eta_d)^{k-1} \ket{k} \bra{k}.
    \label{Eq.DI}
\end{equation}
Here, $\eta_d$ denotes the efficiency of the heralding detector, with $(0 \leq \eta_d \leq 1)$. The mixed nature of input single photon state is taken as a convex combination of the vacuum state and single photon number states \cite{Bimbard2010},
\begin{equation}
    \hat{\rho}_a = (1-\eta_s) \ket{0}\bra{0} + \eta_s \ket{1}\bra{1},
\end{equation}
where $\eta_s$ denotes the non-ideal efficiency of the photon sources. These imperfections alter the pure state of Eq. \ref{Eq.CasDQ} into a mixed state. Thus, the input density operator for the BS is $\hat{\rho}_{in} = \ket{\alpha}\bra{\alpha} \otimes \hat{\rho}_a$. After the BS unitary operation and the effects of the imperfect detector, the resulting non-ideal state is:
\begin{align}
    \hat{\rho}_{R} = & \sum_{k=1}^{\infty}\, k \,\eta_d (1-\eta_d)^{k-1} \bra{k} \hat{U}^{\dagger}_{BS} \hat{\rho}_{in} \hat{U}_{BS}^{} \ket{k}.
    \label{Eq.RQOC}
\end{align}
To evaluate the experimental imperfections in the DQ preparation, it is necessary to measure the overlap between the ideal state and the experimentally realized state. This is achieved by calculating the fidelity $(\mathcal{F}_{R})$, which quantifies the similarity between the ideal density operator $(\hat{\rho}_I=\ket{\psi}_{nm}\bra{\psi})$ and the realized density operator $(\hat{\rho}_R)$. The fidelity is defined as,
\begin{equation}   
    \mathcal{F}_{R} = \hbox{Tr}(\hat{\rho}_{I} \, \hat{\rho}_R).
\end{equation}
Unit fidelity indicates that the ideal and realized states are identical, whereas a value below unity reflects the deviation arising from experimental imperfections. At present, single photon detector efficiencies as high as $98\%$ have been experimentally demonstrated \cite{Peng2020, Chang2021}. Accordingly, the detector efficiency $\eta_d$ is fixed at $0.98$, and the fidelity between the ideal and non-ideal states is evaluated for different levels of source inefficiency.

Figure \ref{Fig.RQOC} illustrates the fidelity between ideal and realistic states, with detector inefficiency fixed at $\eta_d = 0.98$, generated using a cascaded beam splitter setup. The realistic generation of higher Fock states, squeezed states, and LSCS under varying photon source inefficiencies, is depicted in Figs. \ref{Fig.RQOC}(a), (b), and (c), respectively. The fidelity for the experimental generation of Fock states, from lower to higher orders, increases monotonically with increasing source inefficiencies. This trend results from the increasing number of beam splitters, which is directly affected by the source inefficiencies at each input.
A rise and fall pattern in fidelity is observed for squeezed state generation, as shown in Fig. \ref{Fig.RQOC}(b). This behavior arises because the vacuum component in the input sources matches the vacuum component of the generated squeezed states. As previously noted, most finite superposed squeezed states predominantly require a vacuum state component. A similar pattern is observed in Fig. \ref{Fig.RQOC}(c) for the generation of the family of cat states, particularly for even cat states, three headed cat states and the compass state with $h=0$. This pattern is also attributed to the presence of a vacuum state component in these states.
%%%%%%%%%%%%%%%%%%%%%%%%%%%%%%%%%%%%%%%%%%%%%%%%%%%%%%%%%%%%%%%%%%%%%%%%%%%%%%%%%%%%%%%%%%%%%%%%%%%%%%%%%%%
%%%%%%%%%%%%%%%%%%%%%%%%%%%%%%%%%%%%%%%%%%%%%%%%%%%%%%%%%%%%%%%%%%%%%%%%%%%%%%%%%%%%%%%%
{\renewcommand{\arraystretch}{1.25}
\begin{table*}[htbp]
\centering
\caption{Success probability and fidelity of the generated states.}
\begin{adjustbox}{width=0.8\textwidth}
\begin{tabular}{c| c| c| c| c| c}
\hline
DQ & $\abs{\alpha}^2$ & $R_l$ & Created state & $\hbox{S}_{p}$ & Fidelity \\
\hline
\multicolumn{5}{c}{Fock states} \\
\hline
$\ket{\psi}_1$ &  $\frac{1}{1-R}$ & $\frac{\abs{\alpha}^2-1}{\abs{\alpha}^2}$ & $\ket{1}$ & $0.1839$ & 1 \\
$\ket{\psi}_2$ &  5 & 0.50, 0.80 & $\ket{2}$ & 0.0100 & 1 \\
$\ket{\psi}_3$ &  8.65 & 0.5025, 0.7875, 0.8800 & $\ket{3}$ & $2.6988\!\times\!10^{-4}$ & 1 \\
$\ket{\psi}_4$ &  12.575 & 0.5667, 0.8033, 0.8817, 0.9217 & $\ket{4}$ & $6.8629\!\times\!10^{-6}$ & 1 \\
$\ket{\psi}_5$ &  17.85 & 0.4820, 0.7700, 0.8615, 0.9058, 0.9333  & $\ket{5}$ & $2.7505\!\times\!10^{-8}$ & 1 \\
\hline
\multicolumn{5}{c}{Optimal squeezed states} \\
\hline
\multirow{2}{*}{$\ket{\psi}_1$}
    & $\dfrac{1}{4R}\left[\sqrt{3} \pm \sqrt{\dfrac{3+R}{1-R}} \right]^2$ 
    & \multirow{2}{*}{$-$} & \multirow{2}{*}{$\ket{\chi}_1$}  
    & \multirow{2}{*}{0.3388} & \multirow{2}{*}{$0.98$}\\
$\ket{\psi}_2$ &  12.4 & 0.92, 0.49 & $\ket{\chi}_2$ & 0.0022 & 0.98 \\
$\ket{\psi}_3$ &  14.6 & 0.83, 0.95, 0.78 & $\ket{\chi}_3$ & 0.0015 & 0.99 \\
$\ket{\psi}_4$ &  6.55 & 0.80, 0.36, 0.65, 0.12 & $\ket{\chi}_4$ & $5.02\!\times\!10^{-5}$ & 0.98 \\
$\ket{\psi}_5$ &  11.0 & 0.62, 0.25, 0.79, 0.86, 0.27 & $\ket{\chi}_5$ & $6.98\!\times\!10^{-7}$ & 0.97 \\
\hline
\multicolumn{5}{c}{LSCS} \\
\hline
$\ket{\psi}_{4}$ &  9.5 & 0.15, 0.65, 0.82, 0.3 & $\ket{\phi}_{2,0}$ & $5.60\!\times\!10^{-6}$ & 0.995 \\
$\ket{\psi}_{5}$ &  15 & 0.88, 0.43, 0.80, 0.32, 0.40 & $\ket{\phi}_{2,1}$ & $7.67\!\times\!10^{-8}$ & 0.993 \\
$\ket{\psi}_{3}$ &  4.2 & 0.72, 0.46, 0.12  & $\ket{\phi}_{3,0}$ & $9.01\!\times\!10^{-4}$ & 0.994 \\
$\ket{\psi}_{4}$ &  7.5 & 0.88, 0.64, 0.64, 0.56 & $\ket{\phi}_{3,1}$ & $2.94\!\times\!10^{-4}$ & 0.982 \\
$\ket{\psi}_4$ &  12.5 & 0.74, 0.43, 0.82, 0.18  & $\ket{\phi}_{4,0}$ & $4.75\!\times\!10^{-7}$ & 0.995 \\
$\ket{\psi}_{5}$ & 11.2 & 0.88, 0.88, 0.56, 0.56, 0.58 & $\ket{\phi}_{4,1}$ & $5.44\!\times\!10^{-6}$ & 0.982 \\
\hline
\end{tabular}
\end{adjustbox}
\label{tab:FSQLSCS}
\end{table*}
}
%%%%%%%%%%%%%%%%%%%%%%%%%%%%%%%%%%%%%%%%%%%%%%%%%%%%%%%%%%%%%

\section{Summary}
\label{sect.Sum}
In this article, we presented an experimentally feasible optical setup for creating diverse nonclassical states with the current quantum optical technologies. A key advantage is that our cascaded setup surmounts the need for higher Fock states by replacing them with multiple copies of single photon resources and employs single photon detectors. The free parameters to tune the setup are the coherent parameter and reflectivities of beam splitters. Expressing the setup output in displaced qudits format unfolds the target state manipulation and favors tuning the appropriate input parameters. To the end, we strive to bring out the full potential of the setup, which motivates the generation of arbitrary single mode quantum states using our approach.

In recognition, we demonstrated several examples of creating higher photon states by cumulating all the non-Gaussian capacity of single photon sources. Specifically, by using $n$ single photon sources, we can construct displaced $\ket{n}$ photon states with unit fidelity. The realistic preparation of a family of Schrodinger Cat states is reported with fidelities approaching $99\%$. Schrodinger Cat states, three headed cat states, and compass states with different phase rotations are generated for coherent strengths up to $\abs{\gamma}^2 = 2$. In addition, the possible creation of GKP resource states, namely ON states and weak cubic phase states, is shown. All the states are produced with some displacement in the phase space with respect to their input parameters; however, this does not affect the intrinsic nonclassicality. Despite the low success probabilities associated with state generation, fidelities exceeding $0.95$ are consistently achieved, indicating high accuracy in producing the desired states.

Finally, we generalized our methodology and devised a protocol for optimizing the input parameters to get the target states unsophisticated. Also, we accounted for the experimental imperfections by considering the single photon source as a mixture of single photons with vacuum states, along with detector inefficiencies. We analyzed the realistic generation of the target states by fixing detector inefficiency as $0.98$ and for a various range of non-unit single photon sources. Overall, we conclude that the figure of merit used here (displaced qudits) will prompt experimentalists to achieve the desired single mode quantum states using our cascaded setup. 

Having analytical expression for the state representation of the cascaded setup with coherent input embolden our quest for more in depth analysis, for similar construction employing single photon sources and single photon detectors. One of the possible approach is employing squeezed vacuum as input state. It may eliminate the current limitation of the setup studied in this article: low success probability. In addition, the nonclassicality of the squeezed vacuum may also play a vital role in constructing exotic nonclassical states.

\textbf{Data Availability Statement:} \par
This article has no associated data and the data will not be deposited.

\textbf{Acknowledgment:} \par
E. D and A. B. M Ahmed acknowledge the financial support of RUSA phase 2.0 of Madurai Kamaraj University.

\bibliography{apssamp}% Produces the bibliography via BibTeX.
%\end{thebibliography}

\end{document}